\documentclass[journal]{IEEEtran}
\usepackage{bm}
\usepackage{stmaryrd}
\usepackage{array}
\usepackage{cite}
\usepackage{amsmath,amssymb,amsfonts}
\usepackage{algorithm}   
\usepackage{algorithmic}
\usepackage{graphicx}
\usepackage{textcomp}
\usepackage[justification=centering]{caption}
\usepackage{xcolor}
\usepackage{setspace}
\ifCLASSINFOpdf
\else
\fi
\begin{document}
%
\title{VFL: A Verifiable Federated Learning with Privacy-Preserving for Big Data in Industrial IoT}
%
%
%

\author{Anmin Fu, \IEEEmembership{Member, IEEE}, Xianglong Zhang, Naixue Xiong, \IEEEmembership{Senior Member, IEEE},\\ Yansong Gao and Huaqun Wang
\thanks{This work is supported by National Natural Science Foundation of China (61572255, 61941116), the Fundamental Research Funds for the Central Universities (30920021129), Jiangsu Key Laboratory of Big Data Security \& Intelligent Processing, NJUPT(BDSIP1909), and CERNET Innovation Project (NGII20190405). (Corresponding authors: Naixue Xiong).} 
\thanks{A. Fu, X. Zhang and Y. Gao are with the School of Computer Science	and Engineering, Nanjing University of Science and Technology, Nanjing, 210094, PR China. (e-mail: fuam@njust.edu.cn; 1651949523@qq.com; yansong.gao@njust.edu.cn).}
\thanks{N. Xiong is with the Northeastern State University, Department of Mathematics and Computer Science, Tahlequah, OK, USA. (e-mail: xiongnaixue@gmail.com).}
\thanks{H. Wang is with the Jiangsu Key Laboratory of Big Data Security and Intelligent Processing, Nanjing University of Posts and Telecommunications, Nanjing 210023, PR China. (email: whq@njupt.edu.cn).}}

\maketitle

\begin{abstract}
Due to the strong analytical ability of big data, deep learning has been widely applied to model the collected data in industrial IoT. However, for privacy issues, traditional data-gathering centralized learning is not applicable to industrial scenarios sensitive to training sets. Recently, federated learning has received widespread attention, since it trains a model by only relying on gradient aggregation without accessing training sets. But existing researches reveal that the shared gradient still retains the sensitive information of the training set. Even worse, a malicious aggregation server may return forged aggregated gradients. In this paper, we propose the VFL, verifiable federated learning with privacy-preserving for big data in industrial IoT. Specifically, we use Lagrange interpolation to elaborately set interpolation points for verifying the correctness of the aggregated gradients. Compared with existing schemes, the verification overhead of VFL remains constant regardless of the number of participants. Moreover, we employ the blinding technology to protect the privacy of the gradients submitted by the participants. If no more than n-2 of n participants collude with the aggregation server, VFL could guarantee the encrypted gradients of other participants not being inverted. Experimental evaluations corroborate the practical performance of the presented VFL framework with high accuracy and efficiency.
\end{abstract}


%
\IEEEpeerreviewmaketitle

\section{Introduction}
%
%
%
%
\IEEEPARstart{R}{ecently}, the integration of artificial intelligence and industrial IoT has promoted the development of intelligent industry [1], [2], [40]. As a branch of artificial intelligence, due to strong capabilities in modeling, identification, and classification of big data [2], deep learning has been applied to solve data-driven problems in industrial IoT [41]. Nowadays, with the development of industry IoT and the popularity of intelligent products [3], [4], [18], [20], large amounts of data are generated by industry IoT systems and resided in various devices [25], which provides favorable conditions for training high-quality deep learning model. In order to effectively utilize these data, it is a desiderata of exploring deploy deep learning into end devices. Deep learning is hungry about data to increase, in particular, its accuracy, thus it is common practice to collect or share industrial data among different enterprises and institutions. However, these shared industrial data contain sensitive information [5], [16], [39], such as account information, case history, etc. Once these data are shared, clients cannot restrict the purpose for which it is used. For example, in the healthcare system, patients are least willing their diagnoses to be accessible to unauthorized third parties.

Furthermore, there is a trend to strengthen data privacy to prevent misuse of it. The European Union enforced the General Data Protection Regulation (GDPR) on May 25, 2018. It enforces enterprises to employ clarified language in the privacy agreements and reserves clients the right to delete or withdraw their data. In fact, these laws oblige data to be stored in the form of islands [14]. The inability to share data renders in hardness of training an accurate model without harvesting rich data. For medical institutions, each of them is restricted to train a model over the limited local data, such model is with unsatisfied accuracy and may not even be applicable to handle inputs from a different institution that are out of data feature distribution learned by the trained model. Therefore, the privacy issues and legislation put forward new requirements for effective deployment of deep learning to large-scale collaborative industrial applications.

In order to address the industrial data island issue resulted from by the laws and still utilize them legitimately, some secure centralized deep learning schemes have been proposed. In such schemes, the collected data are pre-encrypted [25], or pre-masked with noise [26]. Nevertheless, since the neural network requires nonlinear computation, it is difficult to mount deep learning on encrypted dataset [30], [23]. Training noise-added data is, however, usually accompanied with a trade-off of accuracy drop. Moreover, the computational overhead of secure centralized deep learning makes it extremely challenge to be applicable for large-scale industrial data and deep model structure that are commonly adopted to achieve high accuracy.

In addition to secure centralized deep learning, Google proposed federated learning, a distributed deep learning framework in 2016 [7], [27], [28]. Federated learning only requires participant to share their gradients for aggregation rather than localized data directly. The main benefit is to protect privacy of localized data as they are inaccessible. Compared to centralized deep learning, federated learning avoids the processing of training sets, such as encryption, and improves efficiency. Therefore, federated learning has received widespread attention from both the academia and industry.

Though federated learning greatly reduces privacy leakage surface, it, however, still faces challenges of fundamentally preventing privacy leakage. Phong et al. [9] demonstrated that the aggregation server could approximate the local data by numerical methods. Wang et al. [11] designed the mGAN-AI  (Generative Adversarial Network) model on the aggregation server to recover the original images of target participants. The gradients shared in plaintext is the main reason for these privacy issues. Recently, schemes built upon homomorphic encryption [35] have been proposed, where the gradients are aggregated in the form of ciphertext. But participants use the same secret key, if the aggregation server colludes with any participant to have the key, the ciphertext is no longer meaningful.

Apart from privacy issues, inadvertently, most schemes ignored the verification of the correctness of the aggregated result [6]. Driven by certain illegal interests, the aggregation server may reduce the aggregation operation to save computational cost [12], or forge the aggregated result to impact the model update. Even worse, curious aggregation server could carefully return crafted results to participants for analyzing the statistical characteristics of participant-uploaded data and seduce participants to unintentionally expose more sensitive information [11]. Therefore, the effective verification of the aggregated result shall be always taken into consideration.

In this paper, we propose VFL, a verifiable federated learning with privacy-preserving to achieve efficient and secure model training for industrial intelligent. Specifically, we use Lagrange interpolation and the blinding technology to achieve secure gradient aggregation. Meanwhile, we devise an efficient verification mechanism, where participants can detect forged results with an overwhelming probability. Our contributions are summarized as follows:
\begin{itemize}
	\item
	We propose a verification mechanism for the correctness of the aggregated results based on the characteristics of Lagrange interpolation. In our scheme, each participant can independently and efficiently detect forged results. Compared with existing schemes, the computational overhead of our verification mechanism does not increase with the number of participants. Our verification mechanism also demands for less overhead to each participant.
	\item
	We combines Lagrange interpolation and the blinding technology to realize the secure aggregation of gradients. The joint model and the gradients are protected from the aggregation server. Furthermore, if no more than $n$-2 of $n$ participants collude with the aggregation server, VFL can guarantee that the gradients of other participants will not be leaked.
	\item We provide a comprehensive security analysis for our scheme, which demonstrates the security of the training set and model as well as the verifiability of the VFL framework. We evaluate our scheme over the MNIST dataset with multi-layer perceptron (MLP). The experimental results demonstrate that our scheme has high accuracy and efficiency, and the verification overhead is acceptable to the participants.
\end{itemize}

The rest of the paper is organized as follows: we briefly outline related work in Section~II and introduce preliminaries in Section~III. After that, we present an structure overview of the VFL framework and elaborates on its procedures in Section~IV. The correctness and security analysis for VFL are provided in  Section~V. Experiment evaluation of the proposed scheme is presented in Section VI. Finally, we conclude our paper in Section~VII.

\section{Related Work}
Secure training can be divided into two categories: centralized and distributed. In this section, we briefly describe them accordingly.
\subsection{Secure Centralized Training}
Secure centralized training means the server collects the training sets for training without gaining privacy. To protect the privacy of training sets, Li et al. [36] proposed a multi-key privacy-preserving deep learning in cloud computing, which realizes the conversion of multi-key homomorphic encryption [42] and double decryption mechanism [38] to enable the server trains the model on the ciphertext set. Li et al. [8] implemented naive Bayesian classifiers on multiple data sources using homomorphic encryption and differential privacy. Xu et al. [26] proposed a differential privacy GAN scheme GANobfuscator, which achieves different privacy through GANs by adding carefully designed noise. The scheme generates synthetic data according to the training task instead of using real training data directly. Mohassel et al. [17] proposed the SecureML system based the ABY mixed-protocol [32] and oblivious transfer [34], where participants split their data and respectively sent them to two non-colluding servers to implement the stochastic gradient descent (SGD) algorithm by means of secure multi-party computation. Agrawal et al. [21] utilized key components of DNN training and  improved upon the state-of-the-art in DNN training in two-party computation, which improved the efficiency and accuracy of centralized training.

Secure centralized training requires participant to process the training sets in advance, such as encrypting, adding noise or splitting, and performs training on the processed data. However, due to the complexity of the neural network model, it is inefficient for the server to train the processed data. Therefore, secure centralized training is not suitable for large-scale industrial applications.

\subsection{Secure Distributed Training}
Different from secure centralized training, secure distributed training assigns the training work to participants. Shokri et al. [19] firstly proposed the collaborative deep Learning framework, in which the participants share the gradients and update the model asynchronously. Google extended the framework to the parallel scene and proposed federated learning [7], [27], [28] that requires participants to upload gradients synchronously for aggregation. The two schemes prevent the server being directly access the training sets. Unfortunately, existing researches revealed that this could still lead to privacy breaches [11].

Phone et al. [9] pointed out that the training sets can be recoveried form the gradients by numerical methods, and proposed to employ additively homomorphic encryption to realize secure aggregation. Li et al. [10] used the one-time mask technology to improve upon additively homomorphic encryption. The trainers and the introduced server-aid complete training on the ciphertext set through multiple interactions. Zhou et al. [15] deployed federated learning on the Internet of Things (IoT). The scheme not only protects the data security of IoT devices, but also improves the training efficiency and model accuracy. Abadi et al. [29] proposed a deep learning scheme with differential privacy, in which participants add Laplacian noise to gradients to satisfy differential privacy. Although the above schemes realize the secure gradient aggregation, they do not consider the verification of the aggregated results.

Ma et al. [12] focused on the verifiable federated learning, they take advantage of bilinear aggregate signature [45] to verify the correctness of the aggregated results. Nevertheless, all the participants are required to join the verification process. When the number of participants is large, the verification mechanism results in high cost. Xu et al. [22] adopted the homomorphic hash function and Shamir's secret sharing to realize secure gradient aggregation and verification. The scheme also supports participant drops, but it does not consider protecting the model. The aggregation server can get the final model, but in practice, only participants possess ownership of the model. Zheng et al. [13] proposed the Helen framework, which introduces a secure multi-party computing protocol SPDZ [31] to verify the result and employs multi-key homomorphic encryption to perform secure collaborative training of linear models. Nevertheless, it is specifically designed for linear models and thus dose not support training the models with nonlinear layers. Zhang et al. [37] took advantage of hash function [44] and Paillier encryption [43] to protect the gradients and verify the correctness of the aggregated results. But Paillier encryption brings high cost to participants. And  the verification overhead of the scheme increases with the number of the participants.

Among secure distributed training schemes, few schemes have considered on the verifiability of federated learning. For the existing verifiable schemes, they are with limitations in terms of computational overhead or the type of models supported. Therefore, efficient and verifiable federated learning with non-linear model supporting is still a pressing concern.

\section{Preliminaries}
In this section, we first simply introduce federated learning. Then we give the basic concept of Lagrange interpolation that will be employed to construct our verification mechanism.

\subsection{Federated Learning}\label{AA}
In federated learning [7], each participant and server collaboratively train a unified neural network model. To speed up the convergence of the
model, participants share their gradients to the aggregation server, which aggregates all gradients and returns the result to each participants. The federated learning framework is shown in Fig. 1. Suppose there are $n$ participants, $P_{i}$ $(i=1,2,...,n.)$, who agree on a model architecture. Each round of federated learning can be described as follow.

We denote a neural network model as a function $f(x,M)$, where $x$ is the inputs and $M$ is the model parameter. The model parameter $M$ contains all the biases and the connections between all neurons. Assume that participant $P_{i}$ holds the training set $D^{i}=\left \{ \left \langle x_{j},y_{j} \right \rangle \big|j=1,2,...,T \right \}$, where $x_{j}$ is the input, $y_{j}$ is the label, and $T$ denotes the size of $D_{i}$. The loss function $L_{f}(D^{i},M)$ can be defined as:
\begin{equation*}
L_{f}(D^{i},M)=\frac{1}{T}\sum_{\left \langle x_{j},y_{j} \right \rangle\in D^{i}}(y_{j}-f(x_{j},M))^{2}.
\end{equation*}

The goal of training the neural network model is to find the gradient to update $M$ for minimizing the value of the loss function $L_{f}(D^{i},M)$. In our VFL framework, we adopt the stochastic gradient descent (SGD) to calculate the participant $P_{i}$'s gradient $w_{i}$:
\begin{equation*}
w_{i}=\nabla L_{f}(D^{*i},M),
\end{equation*}
where $\nabla L_{f}$ is derivative of the loss function $L_{f}$. $D^{*i}$ is a random subset of $D^{i}$. Then participants upload their gradients to the aggregation server for aggregation:$w=\sum_{i=1}^{n}w_{i}.$

Finally, the aggregation server sends the aggregated result to each participant and participants update model parameter $M=M-\eta \cdot \frac{w}{n}$ after receiving $w$, where $\eta$ is learning rate. If the termination conditions are not met, then continuing the next round of federated learning.

\begin{figure}[h]
\centering
\includegraphics[width=1.0\linewidth]{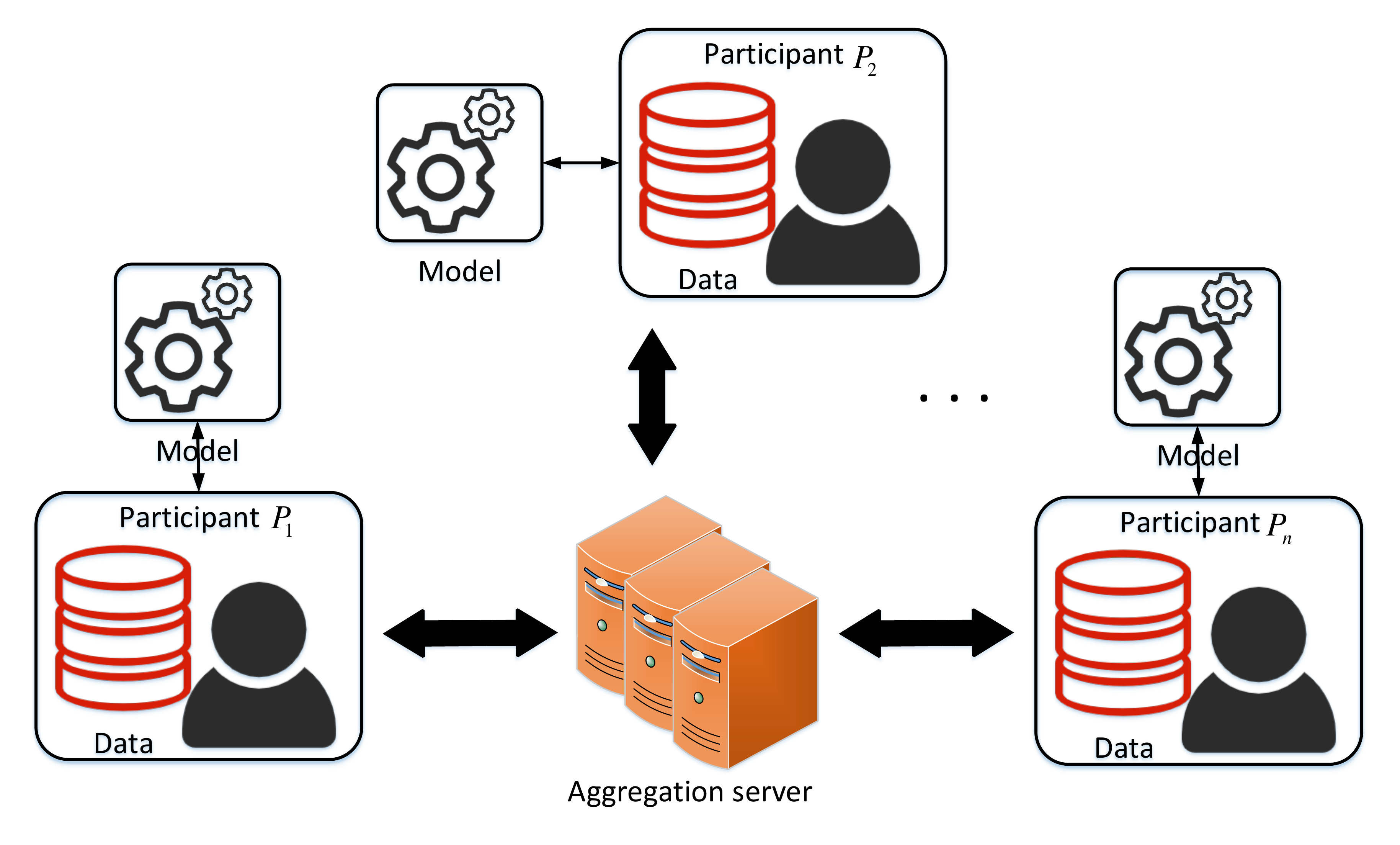}
\caption{\small{The federated learning framework.}}
\label{fig:FLxx}
\end{figure}

\subsection{Lagrange Interpolation}
Lagrange interpolation can find such a polynomial that takes the observed value at each observed point exactly. In the field of cryptography, Lagrange interpolation has been widely applied to secret sharing [33] and coding computing [24], [46]. The following gives a brief introduction to Lagrange interpolation.

\begin{figure*}[h]
	\centering
	\includegraphics[width=1.0\linewidth]{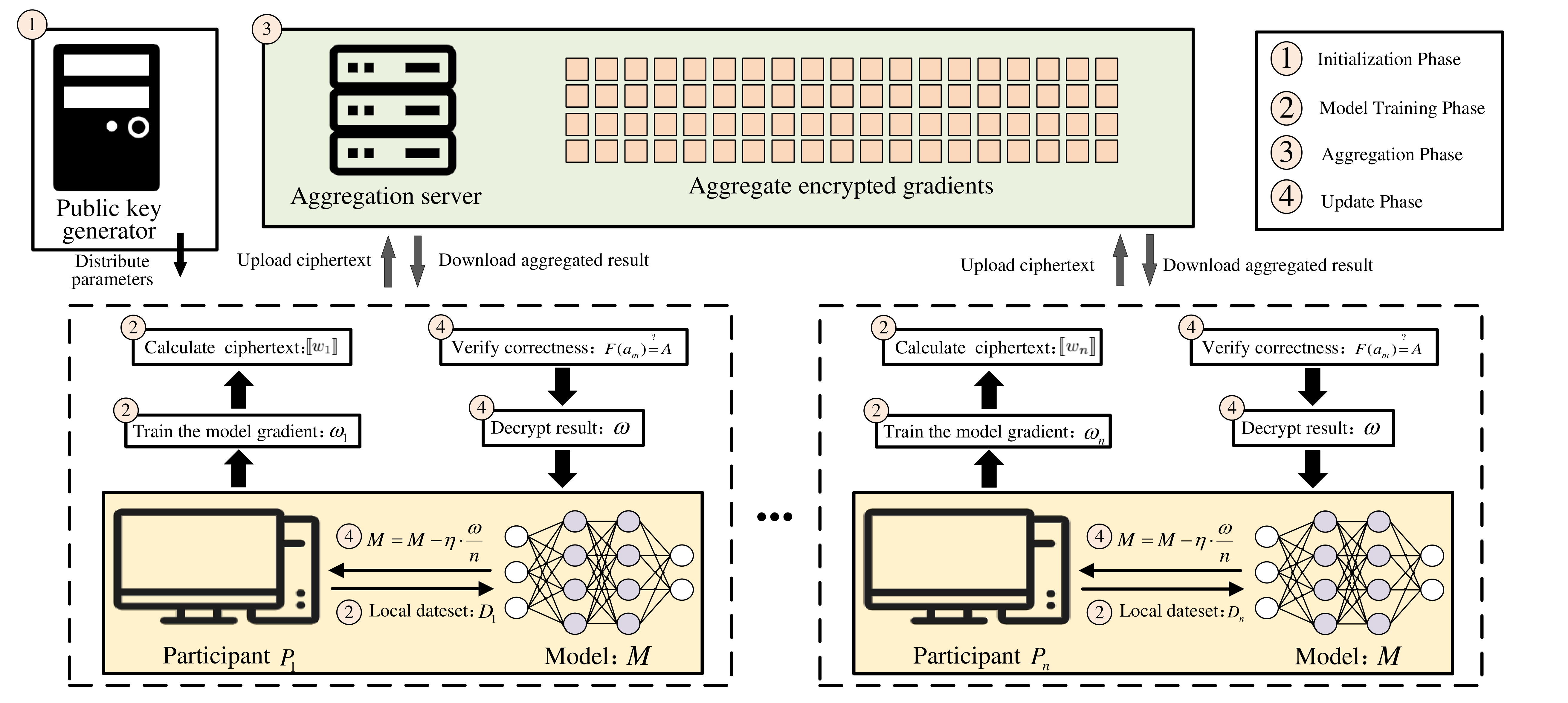}
	\caption{\small{Architecture of the VFL.}}
	\label{fig:1234}
\end{figure*}

Given $n+1$ distinct interpolation points $x_{i}$, $( i=0,1,...,n.)$, together with corresponding numbers $f(x_{i})$, which may or may not be samples of a function $f$, there is a unique $n$-degree polynomial $L_{n}(x)$ satisfying $L_{n}(x_{i})=f(x_{i}),(i=0,1,...,n)$. Let
\begin{equation*}
l_{i}(x)=\prod_{j=0,j\neq i}^{n}\frac{(x-x_{j})}{(x_{i}-x_{j})}, (i=0,1,...,n.).
\end{equation*}
Obviously, $l_{i}(x)$ is also an $n$-degree polynomial and satisfies
\begin{equation*}
l_{i}(x_{j})=\left\{\begin{matrix}
1,j=i\\ 
0,j\neq i\end{matrix}\right.,(i=0,1,...,n.).
\end{equation*}
Then the $n$-degree polynomial $L_{n}(x)$ can be written in Lagrange form:
\begin{equation*}
L_{n}(x)=\sum_{i=0}^{n}f(x_{i})l_{i}(x).
\end{equation*}

Through the above Lagrange interpolation formula, we can conclude that for an $n$-degree polynomial function, selecting any $n$ + 1 points of it can recover the function expression.

\section{Our Proposed VFL Scheme}
This section first gives an overview of the VFL framework, and then details its procedures.

\subsection{Overview}
Fig. 2 presents the architecture of the VFL. The architecture consists of three entities: \textsl{Public Key Generator (PKG)}, \textsl{Participants}, and \textsl{Aggregation Server}.
\begin{itemize}
	\item \textsl{PKG}: The PKG is to initialize the neural network model, generate keys and parameters and distributed them to participants.
	\item \textsl{Participants}: Our scheme is constructed for the IoT and the participant is played by a cluster of end devices: each holds a small amount of data or low diversity of data. The objective of participants is to collaborate in training high-quality neural network models through federated learning. In each round of federated learning, each participant trains the model locally, encrypts their own gradients, and uploads it to the aggregation server. In addition, they receive the aggregated ciphertexts from the aggregation server, verify the correctness of it, and update the model. We assume that the participants are honest and curious, which means they will upload the correct gradient values. However, some participants may collude with the aggregation server in order to obtain the gradient of other participants.
	\item \textsl{Aggregation Server}: In each round of federated learning, the aggregation server aggregates the uploaded ciphertexts, and then distributes the result to each participant. We suppose that the aggregation server is malicious. It may try to steal the privacy information of participants through the received gradients, even worse, forge the aggregated ciphertexts to impact the model update.
\end{itemize}

\begin{figure}[h]
\centering
\includegraphics[width=1.0\linewidth]{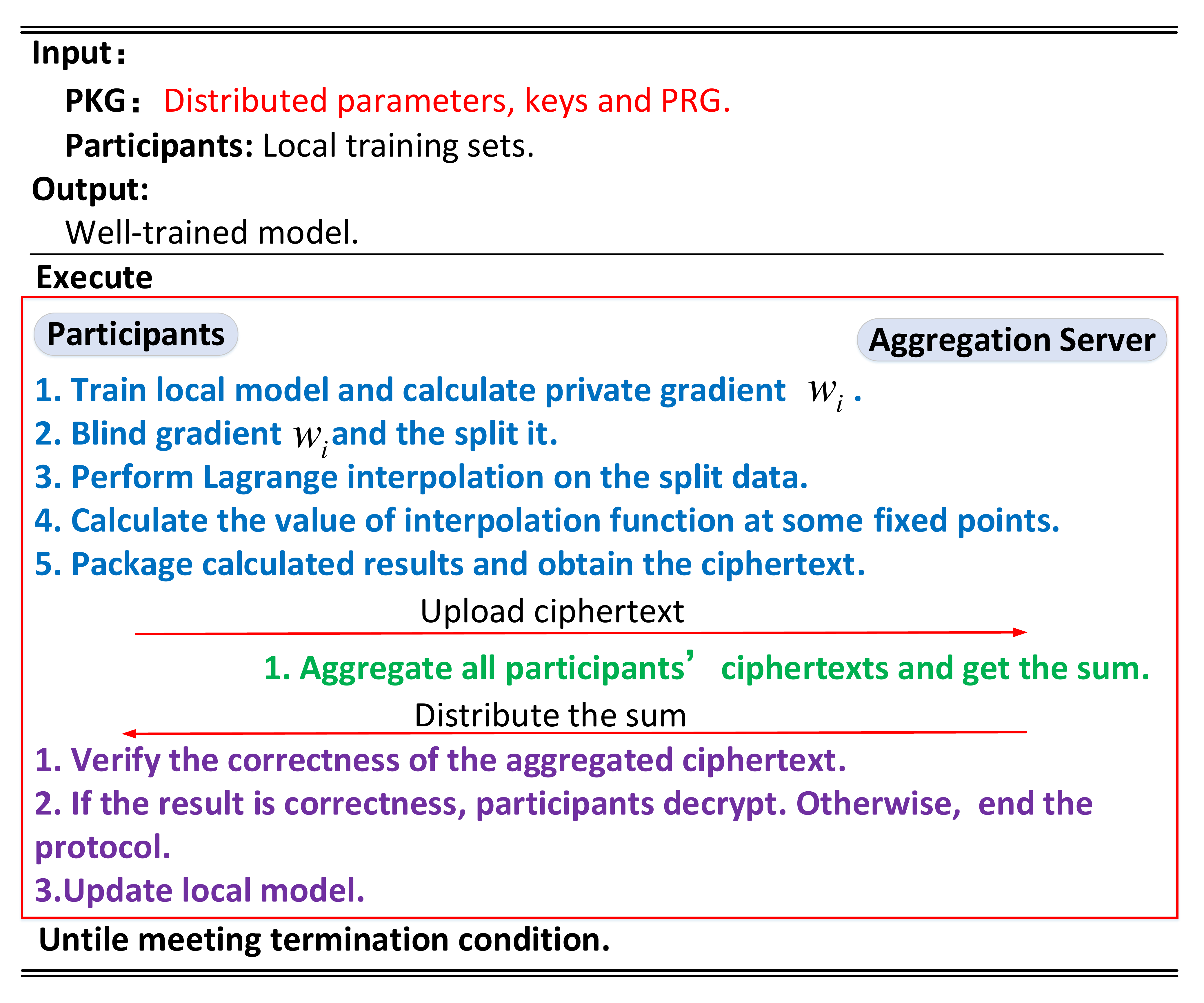}
\caption{\small{The simple summary of different phases.}}
\label{fig:Process}
\end{figure}

As can be seen in Fig. 2, the processes of VFL are divided into four phases: initialization phase, model training phase, model training phase and update phase. We give a simple summary of different phases in Fig. 3, while details are deferred to the following section in the next section. In Fig. 3, red part indicates the initialization phase, blue part represents the model training phase, green part is the aggregation phase and purple part is the update phase.
\begin{itemize}
	\item \textsl{Initialization phase}: The PKG initializes the joint model, generates and distributes keys and parameters to each participant.
	\item \textsl{Model training phase}: The $n$ participants, in parallel, train the model locally and encrypt the calculated gradient, then upload the encrypted gradient to the aggregation server for aggregation.
	\item \textsl{Aggregation phase}: The aggregation server aggregates the received encrypted gradients, and returns the result to all participants.
	\item \textsl{Update phase}: Each participant first verifies the correctness of the result. If the result is correct, they decrypt the aggregated ciphertext and update the model local. Otherwise, abort update.
\end{itemize}

\subsection{Initialization Phase}
The initialization phase is mainly directed by PKG, which generates parameters locally and distributes them to each participant. For simplicity, we assume there are $n$ participants represented by $P_{i}$ $(i \in N)$, where the set $N=\left \{ 1,2,...,n \right \}(n\geq 2)$. In the VFL framework, the PKG needs to generate and distribute the following parameters.

\begin{itemize}
	\item[1)]  According to the neural network architecture agreed by the participants, the PKG randomly generates a learning rate $\eta $ and initializes the model parameter $M$. In addition, all the participants receive $m$ positive integers $g_{1}, g_{2},...,$ $ g_{m}$ that are pairwise co-prime $gcd(g_{i},g_{j})=1$ $(i\neq j)$ and they are large enough to avoid generating overflow error, where $m(m\geq 3)$ is the security parameter of our scheme. We define $G=\prod_{i=1}^{m}g_{i}$. These parameters will be used to package the data.
	\item[2)]  The PKG randomly generates two constant sequences $\left \{ a_{i}|i=1,2,...,m \right \}$ and $\left \{ b_{i}|i=1,2,...,m \right \}$ without intersection. These two constant sequences will be used as interpolation points to process participants' gradients.
	\item[3)]  The PKG sends parameter $A_{i}$ $(i\in N)$ and $A=\sum_{j=1}^{n}A_{j}$ to $P_{i}$ $(i\in N)$, where the specification of parameter $A_{i}$ and $A$ are the same as parameter $M$. In our scheme, the parameter $A$ is used to assist the participants to verify the correctness of the aggregated result.
	\item[4)] The PKG sends each participant two seed sequences and the same pseudo-random generator $PRG(\cdot )$, where the seed sequences received by $P_{i}$ are $\left \{ s_{i}^{(j)}\Big|j\in N,j\neq i \right \}$ and $\left \{ s_{j}^{(i)}\Big|j\in N,j\neq i \right \}$. Obviously, participants $P_{i}$ and $P_{j}$, $(i\neq j)$, have two identical seeds $s_{j}^{(i)}$ and $s_{i}^{(j)}$. This set of parameters will be used to generate pseudo-random numbers to blind participants' gradients.
\end{itemize}

Note that except for parameters $g_{1}, g_{2},...,$ $ g_{m}$ and $G$,  all other parameters and pseudo-random generator are kept secret from the aggregation server.

\subsection{Model Training Phase}
In this phase, each participant $P_{i}$ trains model on the local dataset $D_{i}$ and calculates the private gradient $w_{i}$. In the $t$-th round, participant $P_{i}$ calculates the loss on $D_{i}$'s subset $D^{*i}$.

\begin{equation*}
	L_{f}(D^{*i},M)=\frac{1}{\left | D^{*i} \right |}\sum_{(x_{j},y_{j})\in D^{*i}}(y_{i}-f(x_{i},M)).
\end{equation*}

$P_{i}$ executes back propagation and SGD algorithm to calculate the private gradient $w_{i}=\nabla L_{f}(D^{*i},M)$, and then encrypts it.

\begin{algorithm}[t] 
	\caption{ Gradient encryption algorithm.} 
	\label{alg:Framwork} 
	\begin{algorithmic}[1] 
		\REQUIRE ~~\\ 
		Private gradient $w_{i}$, Pseudo-random generator $PRG(\cdot )$;\\Two seed sequences $\left \{ s_{i}^{j}\right \}$and $\left \{s_{j}^{i}\right \}$, ($j\in N,j\neq i$);\\Two constant sequences $\left \{a_{j}\right \}$ and $\left \{b_{j}\right \}$, ($j=1,2,...,m$).
		\ENSURE ~~\\ 
		Gradient ciphertext $\llbracket w_{i}\rrbracket$;
		\STATE {Blind the gradient $w_{i}:$ \\\quad $w_{i}^{(j),t}\leftarrow PRG(s_{i}^{(j)})^{t}$; $w_{j}^{(i),t}\leftarrow PRG(s_{j}^{(i)})^{t}$;\\\quad
			$\widetilde{w_{i}}\leftarrow w_{i}-\sum_{j=1,j\neq i}^{n}w_{i}^{(j),t}+\sum_{j=1,j\neq i}^{n}w_{j}^{(i),t}$;}
		\STATE Randomly select $m-1$ parameters, which satisfy:\\
		\quad$\widetilde{w_{i}}=\sum_{j=1}^{m-1}v_{i,j}$;
		\label{code:fram:trainbase}
		\STATE Lagrange interpolation is performed on the data set $\left \{(a_{1},v_{i,1}),(a_{2},v_{i,2}),...,(a_{m-1},v_{i,m-1}),(a_{m},A_{i})\right \}$ to obtain the function $F_{i}(x)$:\\\quad
		$F_{i}(x)\leftarrow \sum_{j=1}^{m-1}\left [ v_{i,j}\prod_{k=1,k\neq j}^{m}\frac{(x-a_{k})}{(a_{j}-a_{k})} \right ]+\left [ A_{i}\prod_{k=1}^{m-1}\frac{(x-a_{k})}{(a_{m}-a_{k})} \right ]$;
		\label{code:fram:add}
		\STATE Input $b_{i}(j=1,2,...,m)$ into the function $F_{i}(x)$ and package the results to calculate the ciphertext $\llbracket w_{i}\rrbracket$:\\\quad
		$\llbracket w_{i}\rrbracket\leftarrow CRT\left [ F_{i}(b_{1}),F_{i}(b_{2}),...,F_{i}(b_{m}) \right ];$
		\label{code:fram:classify}
		\STATE Return $\llbracket w_{i}\rrbracket$. 
	\end{algorithmic}
\end{algorithm}

In the VFL framework, the means of post-processing, encrypting, gradients before uploading to the aggregation server is the key to secure gradient aggregation and verifiability. \textbf{Algorithm 1} exemplifies $P_{i}$'s encryption operations. We define $P_{i}$'s private gradient as $w_{i}$, which is the same specification as the model parameter $M$. In round $t$, $P_{i}$ uses the pseudo-random generator $PRG(\cdot )$ to generate two parameter sequences $\left \{ w_{i}^{(j),t}\Big|j\in N,j\neq i \right \}$ and $\left \{ w_{j}^{(i),t}\Big|j\in N,j\neq i \right \}$, which are the same size as $w_{i}$. Define the size of $w_{i}$ to be $\left | w_{i} \right |$. $w_{i}^{(j),t}$ and $w_{j}^{(i),t}$ are composed of the parameters between ($(t-1)\cdot \left | w_{i} \right |)$-th and $(t\cdot \left | w_{i} \right |+1)$-th parameters generated by $PRG(s_{i}^{(j)})$ and $PRG(s_{j}^{(i)})$, respectively. Then $P_{i}$ blinds the gradient $w_{i}$:

\begin{equation}
	\widetilde{w_{i}}=w_{i}-\sum_{j=1,j\neq i}^{n}w_{i}^{(j),t}+\sum_{j=1,j\neq i}^{n}w_{j}^{(i),t}.
\end{equation}

Next, each participant takes advantage of Lagrange interpolation to process the blinded gradient. Firstly, $P_{i}$ splits $\widetilde{w_{i}}$, that is, $P_{i}$ randomly generates $m-1$ parameters $\left \{ v_{i,j},|j=1,2,...,m-1 \right \}$ that satisfies $\widetilde{w_{i}}=\sum_{j=1}^{m-1}v_{i,j}$. Then performing Lagrange interpolation on the data set $\left \{(a_{1},v_{i,1}),(a_{2},v_{i,2}),...,(a_{m-1},v_{i,m-1}),(a_{m},A_{i})\right \}$ to calculate the function $F_{i}(x)$. Notice that the specification of $F_{i}(x)$ is also the same to $w_{i}$, and every element in it is a $(m-1)$-degree polynomial. Then $P_{i}$ feeds the constant sequence $\left \{ b_{i}|i=1,2,...,m \right \}$ into the function $F_{i}(x)$ in turn and gets the result ($F_{i}(b_{1}),F_{i}(b_{2}),...,F_{i}(b_{m})$).

The size of the result is $m$ times that of the original gradient $w_{i}$. In order to reduce the communication overhead, we use the Chinese remainder theorem (CRT) to package the result. The CRT implies that for the following system of congruences:
\begin{equation}
	\begin{matrix}
		y\equiv a_{1}(mod\ g_{1})\\ 
		y\equiv a_{2}(mod\ g_{2})\\ 
		\cdot \cdot \cdot \\
		y\equiv a_{m}(mod\ g_{m})\end{matrix},
\end{equation}
there is a unique solution $y\equiv (a_{1}H_{1}G_{1}+a_{2}H_{2}G_{2}+\cdot \cdot \cdot +a_{m}H_{m}G_{m})mod\ G$ in the finite field $\mathbb{F}_{G}$, where the $G_{i}=G/g_{i}$, and $H_{i}=G_{i}^{-1}mod\ g_{i}$ is the inverse element of $G_{i}$ of module $g_{i}$. Therefore, the data set $(a_{1}, a_{2},..., a_{m})$ can be represented as $y$. To ease description, we write the packaged data $y$ as $CRT\left [ a_{1}, a_{2},...,a_{m} \right ]$. 

In our scheme, the participant $P_{i}$ performs the CRT on the result ($F_{i}(b_{1}),F_{i}(b_{2}),...,F_{i}(b_{m})$) corresponding to modules $g_{j}$ $(j=$$1,2,...,m)$ and get the packaged data $\llbracket w_{i}\rrbracket$ = $CRT\left [ F_{i}(b_{1}) ,F_{i}(b_{2}),...,F_{i}(b_{m}) \right ]$, which is the ciphertext of gradient $w_{i}$. Obviously, due to packaging, the size of $\llbracket w_{i}\rrbracket$ is the same as $w_{i}$. Thus our scheme has the same communication overhead as the original federated learning [7].
Finally, participants send their ciphertexts to the aggregation server.

\subsection{Aggregation Phase}
After the aggregation server receives the ciphertexts uploaded by all the participants, it performs aggregation. The calculation of this process is $\llbracket w\rrbracket=\sum_{i=1}^{n}\llbracket w_{i}\rrbracket$. For two packaged data $y^{(1)}= CRT[a^{(1)}_{1}, a^{(1)}_{2},..., a^{(1)}_{m}]$ and $y^{(2)}=CRT[a^{(2)}_{1}, a^{(2)}_{2},..., a^{(2)}_{m}]$ corresponding to modulus $g_{1}, g_{2}, ..., g_{m}$, the following equation holds:
\begin{equation}
	y^{(1)}+y^{(2)}\equiv a^{(1)}_{i}+a^{(2)}_{i}(mod\  g_{i}), (i=1,2,...,m)
\end{equation}

 The equation (3) implies the CRT satisfies additive homomorphism, so the equation (4) holds.
\begin{equation}
	\begin{split}
		\llbracket w\rrbracket&=\sum_{i=1}^{n}\llbracket w_{i}\rrbracket\\
		&=CRT\left [ \sum_{i=1}^{n}F_{i}(b_{1}), \sum_{i=1}^{n}F_{i}(b_{2}),..., \sum_{i=1}^{n}F_{i}(b_{m}) \right ]\\
		&=CRT\left [ F(b_{1}), F(b_{2}),..., F(b_{m}) \right ],
	\end{split}
\end{equation}	
where $\llbracket w\rrbracket$ is the aggregated ciphertext, and $F(x)=\sum_{i=1}^{n}F_{i}(x)$. Every element in $F(x)$ is also a $(m-1)$-degree polynomial. Note that the aggregation server can not acquire the expression of the function $F(x)$. Because the
constant sequence $\left \{ b_{i}|i=1,2,...,m \right \}$ is kept secret from the aggregate server, which only has the knowledge of the values of
the function $F(x)$ at some unknown points. Afterwards, the aggregation server returns $\llbracket w\rrbracket$ to each participant.

In aggregation phase, driven by certain illegal interests, a malicious aggregation server may forge the aggregated result to impact the model update or attempt to learn sensitive information of the private gradient. Participants should be given the ability to verify the correctness of the results and detect such malicious behavior.

\subsection{Update Phase}
After receiving the aggregation value $\llbracket w\rrbracket$, each participant firstly unpacks and then verifies the correctness of it. Unless the aggregation value is correct, the participant decrypts $\llbracket w\rrbracket$ and updates the local model. \textbf{Algorithm 2} gives the verification and decryption algorithm of our scheme.

Firstly, each participant unpacks the ciphertext $\llbracket w\rrbracket$ by modular operation and then conducts the Lagrange interpolation on the data set $\left \{(b_{1},F(b_{1})),(b_{2},F(b_{2})),...,(b_{m},F(b_{m}))\right \}$ to calculate the expression of the function $F(x)$:
\begin{equation}
	F(x)=\sum_{i=1}^{m}F(b_{i})l_{i}(x),
\end{equation}
where $l_{i}(x)=\prod_{j=1,j\neq i}^{m}\frac{(x-b_{j})}{(b_{i}-b_{j})}$.
Then participants input $x=a_{m}$ into the function and calculate $F(a_{m})$. If the equation $\bm{F(a_{m})=A}$ holds, the aggregation value is correct; otherwise, it is deemed to be a forged result and federated learning ends. Note that the size of the $F(x)$ is the same as the model parameter $M$, and each element in it is one variable function, so the process is actually the verification of multiple equations.

\begin{algorithm}[t] 
	\caption{ Verification and decryption algorithm.} 
	\label{alg:Framwork} 
	\begin{algorithmic}[1] 
		\REQUIRE ~~\\ 
		Ciphertext $\llbracket w\rrbracket$;\\ two constant sequences $\left \{a_{j}\right \}$ and $\left \{b_{j}\right \}$, ($j=1,2,...,m$);
		\ENSURE ~~\\ 
		Plaintext $w$;
		\STATE Unpack ciphertext $\llbracket w\rrbracket$;\\\quad
		$F(b_{i})\leftarrow \llbracket w\rrbracket\ mod\ g_{i}, (i=1,2,...,m);$
		\STATE {Lagrange interpolation is performed on the data set $\left \{(b_{1},F(b_{1})),(b_{2},F(b_{2})),...,(b_{m},F(b_{m}))\right \}$ to obtain the function $F(x)$}:\\\quad
		$F(x)\leftarrow \sum_{j=1}^{m}\left [ F(b_{j})\prod_{k=1,k\neq j}^{m}\frac{(x-b_{k})}{(b_{j}-b_{k})} \right ]$;
		\label{ code:fram:extract }
		\STATE Calculate $F(a_{m})$ to verify the correctness of the aggregated result:\\
		if {$\bm{F(a_{m})}=A$} then\\\quad
		Calculate $F(a_{1}), F(a_{2}),..., F(a_{m-1})$ to decrypt $\llbracket w\rrbracket$: \\\quad \quad
		$w\leftarrow \sum_{j=1}^{m-1}F(a_{j})$;\\
		else\\\quad
		End the protocol;
		\label{code:fram:classify}
		\STATE Return $w$. 
	\end{algorithmic}
\end{algorithm}

 If the aggregated result is correct, each participant inputs the constant sequence $\left \{ a_{i}|i=1,2,...,m-1 \right \}$ into $F(x)$ in turn, and sum up the output results. In this case, the aggregated value $w$ of participants' original gradients is
\begin{equation}
	w=\sum_{i=1}^{n}w_{i}=\sum_{j=1}^{m-1}F(a_{j}).
\end{equation}

At the end of this phase, each participant updates the model parameter $M$ locally: $M= M-\eta \cdot \frac{w}{n}$. After this, the next round of federated learning will be performed until the termination condition is met.

\section{Effectiveness Analysis for VFL}
This section, we give a theoretical analysis of the VFL framework in terms of correctness, data privacy, and verifiability.
\subsection{Correctness}
\textbf{Theorem 1.} \textsl{In the VFL framework, if each entity executes the protocol honestly, participants can obtain correct aggregated gradients to update the model.}

\textsl{Proof:} If each entity executes the protocol honestly in VFL, the correct aggregated gradients can be obtained by participants only if equation (6) holds. Next, we prove that equation (6) holds. According to Algorithm 1,
\begin{equation}
\begin{split}
\sum_{j=1}^{m-1}F(a_{j})&=\sum_{j=1}^{m-1}(\sum_{i=1}^{n}(F_{i}(a_{j})))\\
&=\sum_{i=1}^{n}(\sum_{j=1}^{m-1}(F_{i}(a_{j})))\\
&=\sum_{i=1}^{n}(\sum_{j=1}^{m-1}v_{i,j})\\
&=\sum_{i=1}^{n}\widetilde{w_{i}}.
\end{split}
\end{equation}

In addition,
\begin{equation}
\begin{split}
\sum_{i=1}^{n}(\sum_{j=1,j\neq i}^{n}(w_{i}^{(j),t}))&=\sum_{i=1}^{n}(\sum_{j=1}^{n}w_{i}^{(j),t})-\sum_{i=1}^{n}w_{i}^{(i),t}\\
&=\sum_{i=1}^{n}(\sum_{j=1,j\neq i}^{n}w_{j}^{(i),t}).
\end{split}
\end{equation}    
According to equation (1),
\begin{equation}
\begin{split}
\sum_{i=1}^{n}\widetilde{w_{i}}&=\sum_{i=1}^{n}w_{i}-\sum_{i=1}^{n}(\sum_{j=1,j\neq i}^{m-1}(w_{i}^{(j),t}))+\sum_{i=1}^{n}(\sum_{j=1,j\neq i}^{m-1}(w_{j}^{(i),t}))\\
&=\sum_{i=1}^{n}w_{i}=w.
\end{split}
\end{equation}    

From the above three equations, we can conclude that equation (6) holds. Hence, if each entity in the VFL framework honestly executes the protocol, the participant can obtain the correct aggregated gradients to update the model.
\subsection{Data Privacy}
The gradients retain sensitive information about the training set, the aggregation server can steal participants' privacy from the gradients [9], [11]. The final model parameter $M$ is the `wealth' of participants. Thus, we aim to protect the private gradients and the model parameter in our scheme. 

\textbf{Theorem 2.} \textsl{In the VFL framework, each participant's private gradient $w_{i}$ and the model parameter $M$ will not be leaked to the aggregation server.}

\textsl{Proof:} Protecting model parameter $M$ needs to make sure that the aggregation value $w$ is not leaked in each round. In the VFL framework,
if the aggregation server tries to calculate the private gradient $w_{i}$ and the aggregation value $w$, it needs to get $\left \{ F_{i}(a_{1}),F_{i}(a_{2}),...,F_{i}(a_{m-1}) \right \}$ and $\left \{ F(a_{1}),F(a_{2}),...,F(a_{m-1}) \right \}$, respectively. However, the constant sequence $\left \{ b_{i}|i=1,2,...,m \right \}$ is kept secret from the aggregation server, so the aggregation server cannot calculate the expression of function $F_{i}(x)$ and $F(x)$ through $\left [ F_{i}(b_{1}),F_{i}(b_{2}),...,F_{i}(b_{m}) \right ]$ and $\left [ F(b_{1}),F(b_{2}),...,F(b_{m}) \right ]$ . In addition, the constant sequence $\left \{ a_{i}|i=1,2,...,m-1 \right \}$ is also kept secret from the aggregation server.

In the finite field $\mathbb{F}_{q}$, the probability that the aggregation server obtains the two constant sequences is $\frac{1}{C_{q}^{m}\times C_{q-m}^{m-1}}$, where $C_{q}^{m}$ is the combinatorial number and $C_{q}^{m}=\frac{q!}{m!(q-m)!}$. Consequently, the larger the value of $C_{q}^{m}\times C_{q-m}^{m-1}$ is, the more secure $w_{i}$ and $w$ are. Moreover, the value of $q$ is always very large, such as $2^{64}$. Hence, in the VFL framework, the private gradient $w_{i}$ and the model parameter $M$ will not be leaked to the aggregation server. 

\textbf{Theorem 3.} \textsl{In the VFL framework, if the aggregation server colludes with $k(k \leq n-2)$ participants, the private gradients of other participants will not be leaked.}

\textsl{Proof:} Without loss of generality, we assume that participants $P_{1}, P_{2}, ...,$ and $P_{k}$$(k \leq n-2)$ collude with the aggregation server. Note that the seeds $s_{i}^{j}$ and $s_{j}^{i}$, $(i,j>k)$, are kept secret from them. Accroding to equation (1), from the data uploaded by other participants, they can only get:

\begin{equation}
\begin{split}
\widetilde{w_{i}}+\sum_{j=1}^{k}w_{i}^{(j),t}-\sum_{j=1}^{k}w_{j}^{(i),t}=
w_{i}-\sum_{j= k+1,j\neq i}^{n}w_{i}^{(j),t}+\\\sum_{j= k+1,j\neq i}^{n}w_{j}^{(i),t},(i>k,i \in N).
\end{split}
\end{equation}

The private gradient $w_{i}$ is blinded by $-\sum_{j= k+1,j\neq i}^{n}w_{i}^{(j),t}+\sum_{j= k+1,j\neq i}^{n}w_{j}^{(i),t}$, they cannot obtain the private gradients of other participants. Hence, if the aggregation server colludes with $k(k \leq n-2)$ participants, others' gradients will not be subject to privacy threats.

\subsection{Verifiability}
Driven by certain illegal interests, the aggregation
server may reduce the aggregation operation to save computational cost [12], or forge the aggregated result to impact the
model update. The VFL gives participants the ability to verify the correctness of aggregated results.

\textbf{Theorem 4.} \textsl{In the VFL framework, each participant can independently verify the correctness of the aggregated result, and the verification mechanism can detect forged results with an overwhelming probability.}

\textsl{Proof:} If participants receive the correct aggregated result $\left [ F(b_{1}), F(b_{2}),..., F(b_{m}) \right ]$, obviously, $F(x)$ satisfies the following condition:
\begin{equation}
F(a_{m})=\sum_{i=1}^{n}F_{i}(a_{m})=\sum_{i=1}^{n}A_{i}=A.
\end{equation}
Each participant can check whether equation (11) holds locally, hence, they can verify the correctness of the aggregated result independently. 

When participants receive a forged aggregated result, without loss of generality, we suppose the aggregation server tampered with the aggregated result to 
\begin{equation}
\left [ F(b_{1})+\Delta x_{1},F(b_{2})+\Delta x_{2},...,F(b_{m})+\Delta x_{m} \right ],
\end{equation}
where $\Delta x_{i}$ is the variation of $F(b_{i})$ and $\sum_{i=1}^{m}(\Delta x_{i})^{2}\neq 0$. If the aggregation server attempts to evade the verification mechanism of our scheme successfully, it needs to ensure that the following equation holds:
\begin{equation}
F^{\ast }(a_{m})=\sum_{i=1}^{n}\left [ (F(b_{i})+\Delta x_{i})\cdot l_{i}(a_{m}) \right ]=A,
\end{equation}
where $F^{\ast }(x)$ is calculated from the forged result and $l_{i}(x)$ is given in the equation (5). It can be proved from the equation (5) and (11) that equation (13) is equivalent to 
\begin{equation}
\sum_{i=1}^{n}\left [ \Delta x_{i}\cdot l_{i}(a_{m}) \right ]=0.
\end{equation}

Therefore, the aggregation server only needs to make the equation (14) hold. However, $l_{i}(a_{m})$ is related to the sequence $\left \{ b_{i}|i=1,2,...,m \right \}$ and the constant $a_{m}$, they are confidential to the aggregation server. In the finite field $\mathbb{F}_{q}$, the aggregation server takes these parameters with a probability of $\frac{1}{C_{q}^{m+1}}$. Thus, $l_{i}(a_{m})$ is kept secret from the aggregation server. Therefore, it is impossible for the aggregation server to forge a result to make the equation (13) hold.

From the above, we can conclude that the verification mechanism of our scheme can effectively verify the correctness of the aggregated result.
\section{Performance Analysis}
In this section, we evaluate the performance of VFL framework from model accuracy, computational and communication overhead, by conducting extensive experiments on representative dataset.

\subsection{Experimental Setup}
Our simulation experiment is conducted on Intel(R) Core(TM) i5-9400, 2.90 GHz, and 16 GB memory. We use PC to simulate participants and Alibaba Cloud as the aggregation server. Our codes are in python. The experiments are organized on the MNIST image dataset. MNIST dataset is composed of 70,000 images of 28$\times$28 pixels, handwritten digital grayscale images with labels, among which 60,000 are training data, and 10,000 are test data. We use a popular neural network: multi-layer perceptron (MLP) as the trained model for federated learning. The learning goal is to classify the input into 0-9 possible numbers. 

\begin{figure}[h]
	\centering
	\includegraphics[width=0.9\linewidth]{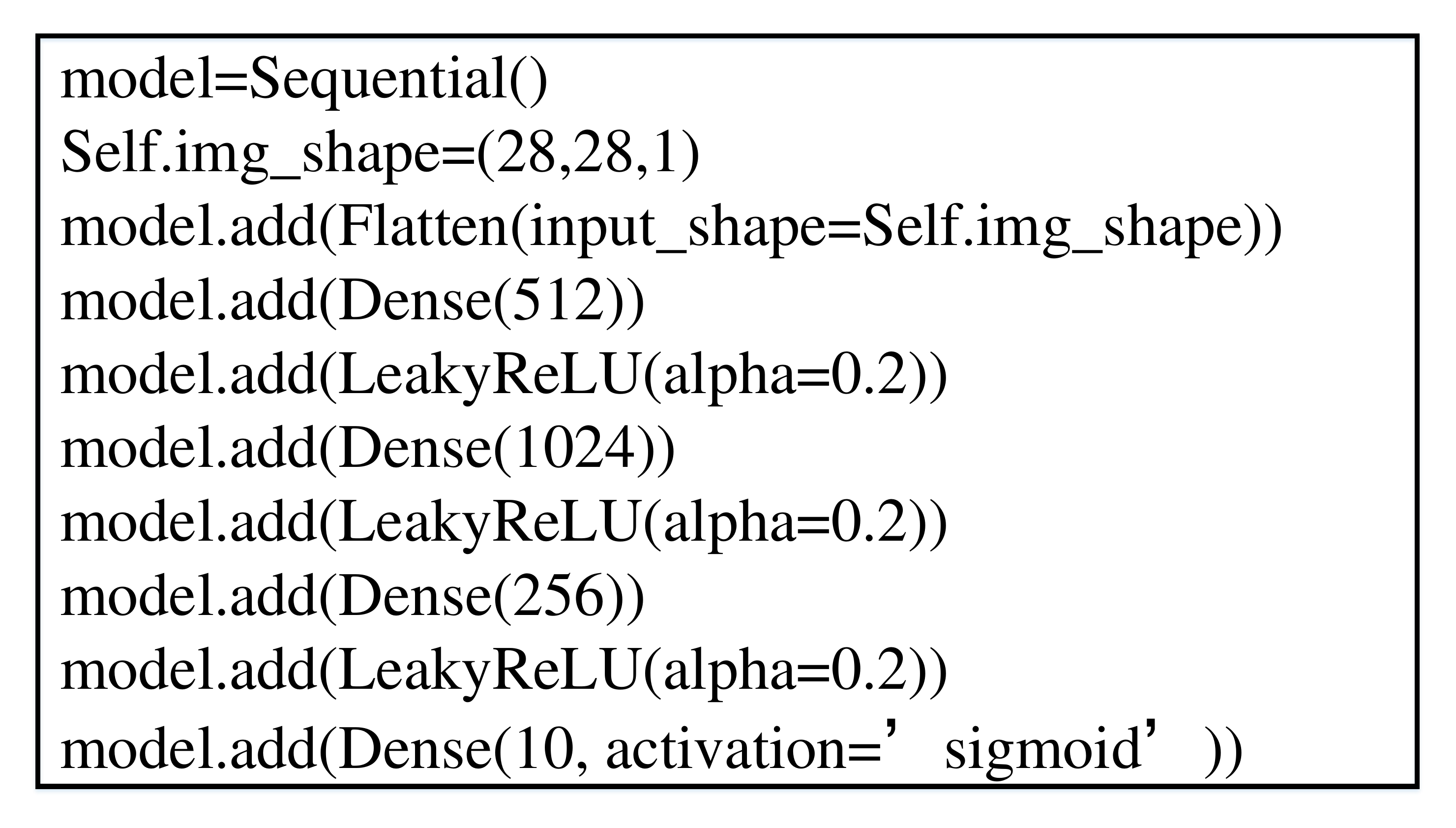}
	\caption{\small{Multi-layer perceptron with 784 inputs.}}
	\label{fig:jiagou}
\end{figure}

Fig. 4 shows the architecture of the MLP, which is 784(input)-512(hidden)-1024(hidden)-256(hidden)-10(output). The number of gradients for the MLP architecture is (784+1)$\times$ (512)+(512+1)$\times$1024+(1024+1)$\times$256+(256+1)$\times$10=1192202. We denote each image as a column vector by reading the image gray value column by column and normalize the gray value vector as the input of the MLP. The learning rate we set is $10^{-2}$, and learning rate decay is $10^{-5}$. In our experiments, we set the computing precision is 64-\textsl{bit} and the number of participants is 20.

\subsection{Model Accuracy}
 In this experiment, we set $m=4$. In fact, we have proved in \textbf{Theorem 1} that no matter what the value of $m$ is, the participants can get the correct aggregated gradient to update the model in VFL, so the $m$ does not affect the accuracy of the model. Fig. 5 shows the accuracy of our VFL and the original federated learning scheme [7] under different rounds. After 400 rounds of training, the accuracy of the VFL framework training model can reach about 94\%, and that of the scheme [7] is approximately 95\%. Experiment results confirm that the VFL framework hardly sacrifices the model accuracy to protect data privacy. 
\begin{figure}[h]
	\centering
	\includegraphics[width=1.0\linewidth]{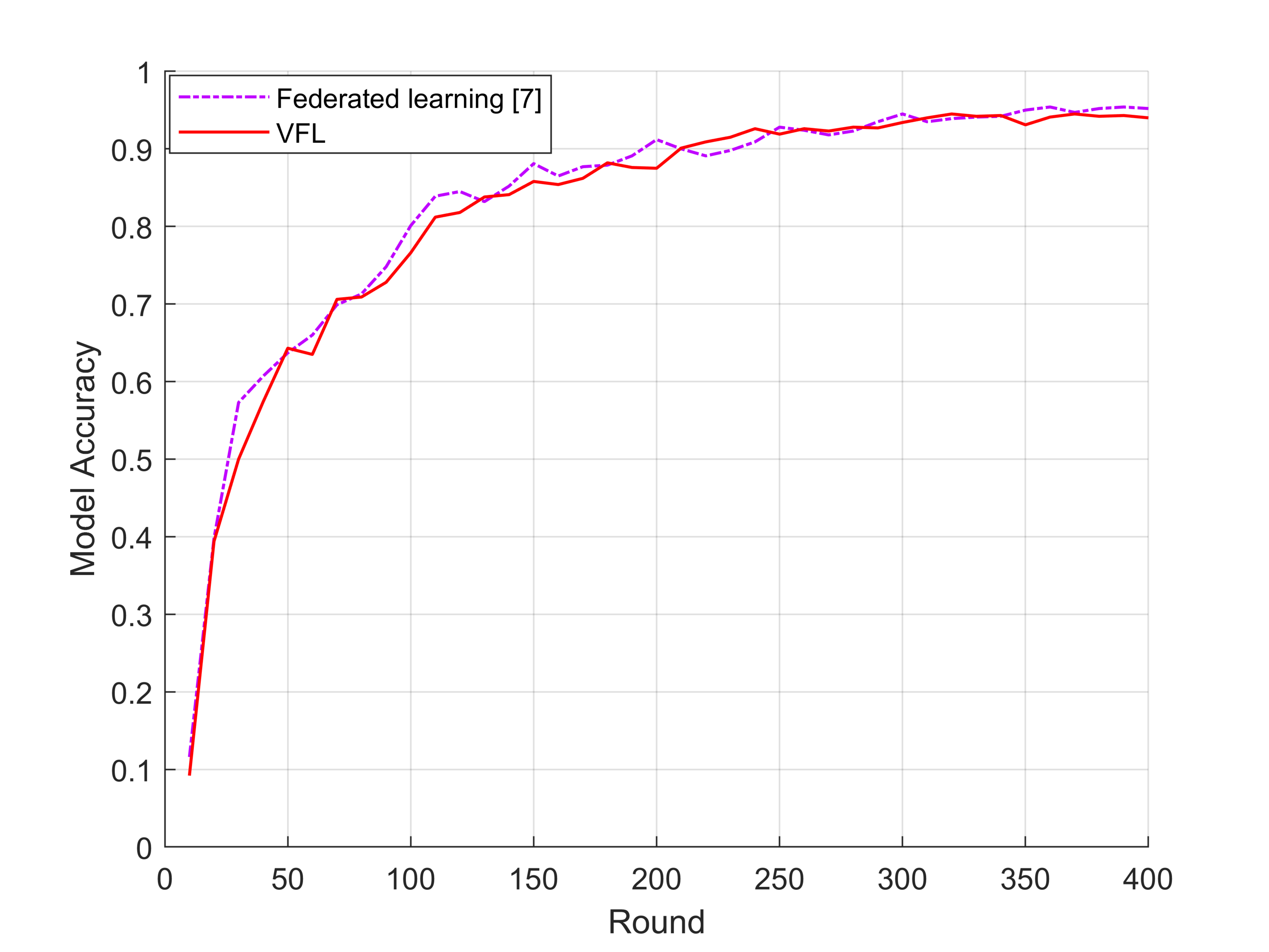}
	\caption{\small{Accuracy comparison between our scheme and Scheme [7]}}
	\label{fig:untitled(6)}
\end{figure}

\subsection{Computational Overhead}
This section evaluates the computational overhead of the VFL framework from three aspects: encryption, decryption, and verification, which proves that compared with the exiting schemes, our scheme brings less computational cost to participants. We compare our scheme with the existing Phone's scheme [9] and Ma's scheme [12]. Since the VFL framework's encryption algorithm is related to the security parameter $m$, we make experimental evaluations of the scheme in two cases: $m$=4 and $m$=8. In addition, we set the size of all secret keys to 64-\textsl{bit}. In this case, even if $m$ is 4, the probability of the aggregation server stealing data privacy is less than $\frac{1}{(C_{2^{64}-3}^{3})^{2}}\ll 1.9\times 10^{-53}$ in the VFL, where $C_{2^{64}-3}^{3}$ is a combinatorial number.

\textsl{1) Computational Overhead of Encryption}

In each scheme, the participant trains the given multi-layer perceptron model, and then encrypts private gradient parameters. Fig. 6 compares the gradient encryption overhead of each participant in different schemes. The experimental results show that in all schemes, the encryption overhead of each participant increases linearly with the number of gradient parameters. For the 1192202 gradient parameters in the given MLP, under different $m$, the encryption overhead of our scheme are $T_{e,m=4}=1.383$s and $T_{e,m=8}=4.916$s, respectively. Since the larger $m$ is, the more interpolation datasets are required for encryption, the higher the overhead is. The increased security reduces the efficiency of the VFL. In addition, the encryption overhead of Phone's scheme [9] (based on LWE encryption) and Ma's scheme [12] (based on ElGamal encryption) are 8.178s and 19.741s, respectively. However, because all participants hold the same key, scheme [9] and [12] cannot resist the collusion attack. Therefore, our scheme has advantages in terms of encryption overhead and privacy protection. 
\begin{figure}[h]
	\centering
	\includegraphics[width=1.0\linewidth]{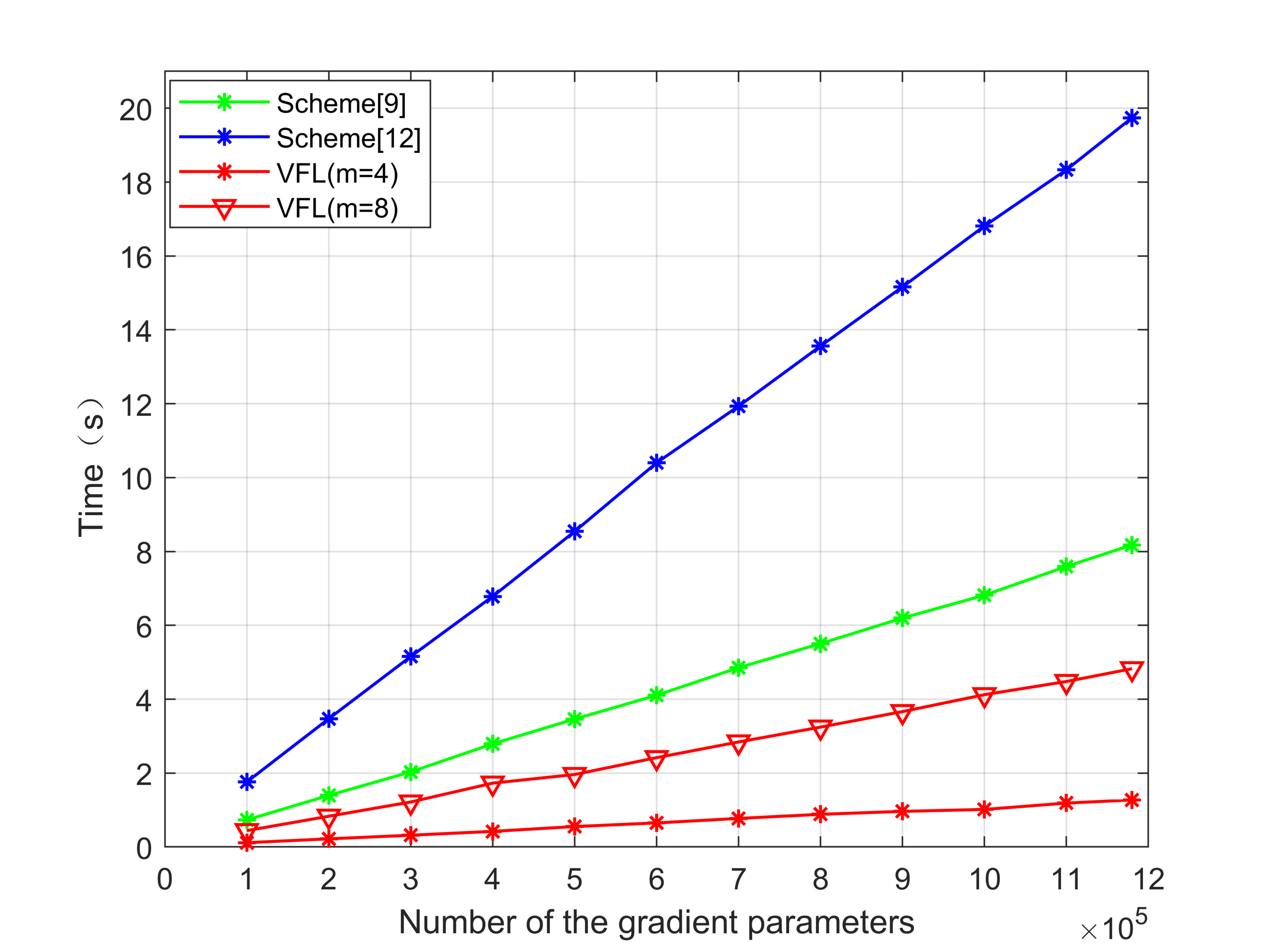}
	\caption{\small{Encryption overhead of each participant in different schemes.}}
	\label{fig:untitled1}
\end{figure}

\begin{figure}[t]
	\centering
	\includegraphics[width=1.0\linewidth]{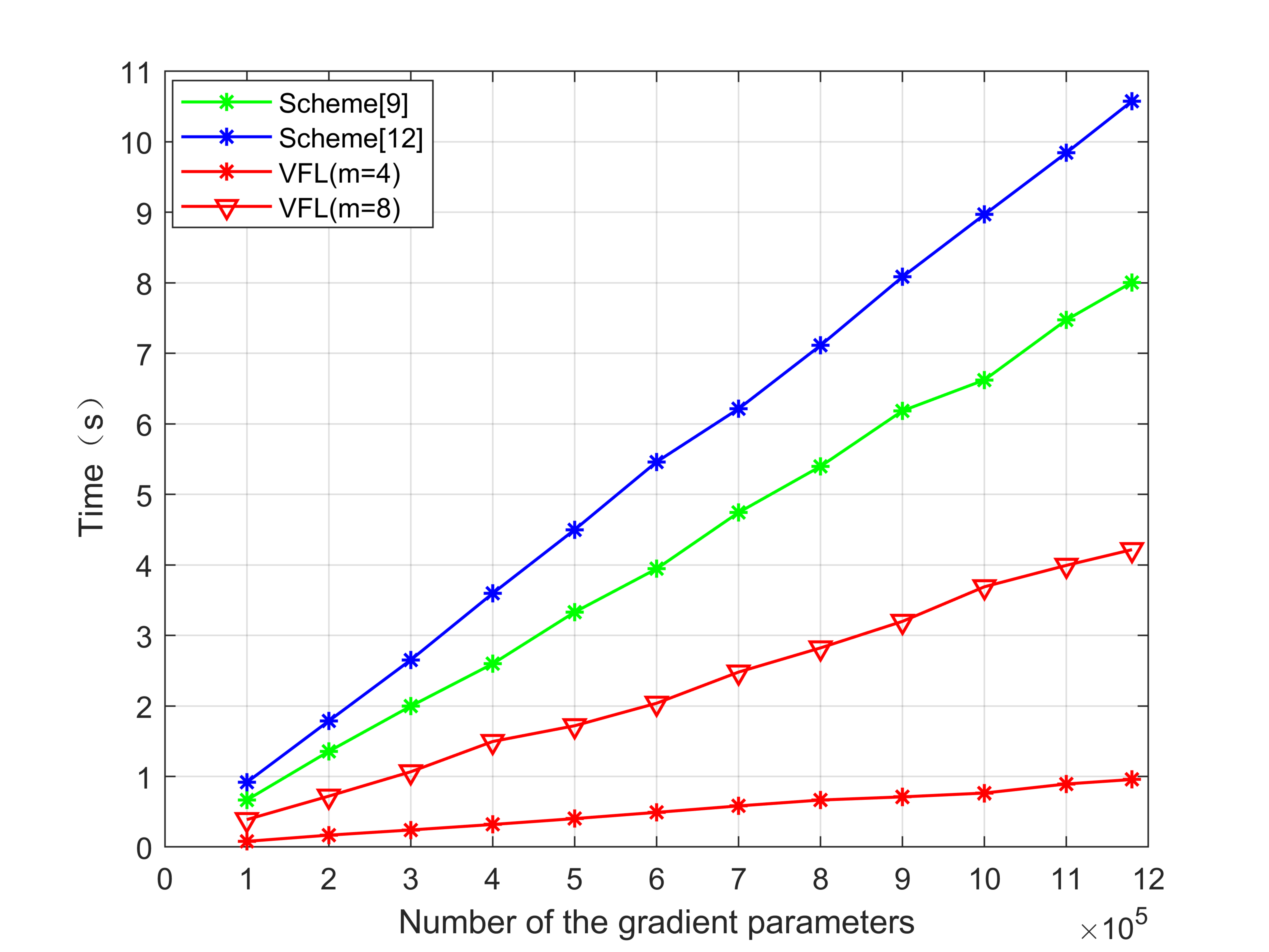}
	\caption{\small{Decryption overhead of each participant in different schemes}}
	\label{fig:untitled234567}
\end{figure}
\textsl{2) Computational Overhead of Decryption}

Fig. 7 shows the decryption overhead of each participant in different schemes. For the MLP model, the decryption overhead of our scheme are $T_{d,m=4}=0.961$s and $T_{d,m=8}=4.379$s, respectively. In our scheme, when the participant decrypts, $m-1$ interpolation points need to be input, the decryption process can be seen as the addition of $m$ equations. In addition, the decryption time cost of the scheme [9] is 8.007s and the scheme [12] is 10.577s. We can conclude that even when $m=8$, the decryption overhead of scheme [9] and scheme [12] is still nearly twice that of our scheme. Consequently, our scheme has high efficiency in decryption.

\textsl{3) Computational Overhead of Verification}

Fig. 8 shows the experimental results of the verification overhead in different schemes. Since the scheme [9] does not support the verification, it is not evaluated for comparison. We set up the case of 10 groups of participants and compared the verification overhead of different schemes in each case. The results confirm that with the increase in the number of participants, the verification overhead of the scheme [12] increases linearly, while our scheme remains constant. When the number of participants reaches 20, the verification overhead of our scheme are $T_{v,m=4}=0.325$s and $T_{v,m=8}=0.623$s respectively, while the scheme [12] is up to 14.624s. 

The equation (5) indicates that the computational overhead of $F(a_{m})$ is related to $l_{i}(a_{m})(i=1,2,...,m)$. The calculation of function $l_{i}(x)(i=1,2,...,m)$ is related to the parameter $m$ and independent on the number of participants $n$. Therefore, the verification overhead of our scheme is insensitive to the number of participants. The scheme [12] employs  bilinear aggregate signature to verify correctness. But all the participants are required to participate in the verification process. Therefore, our verification mechanism is more flexible. In addition, The scheme [12] requires that when the participants receive the aggregated result, they should firstly decrypt and then verify. If the aggregated result is forged, this would waste participants' computing resources. Such a cumbersome is eschewed in the VFL which, in contrast, verifies first before decryption.
\begin{figure}[t]
	\centering
	\includegraphics[width=1.0\linewidth]{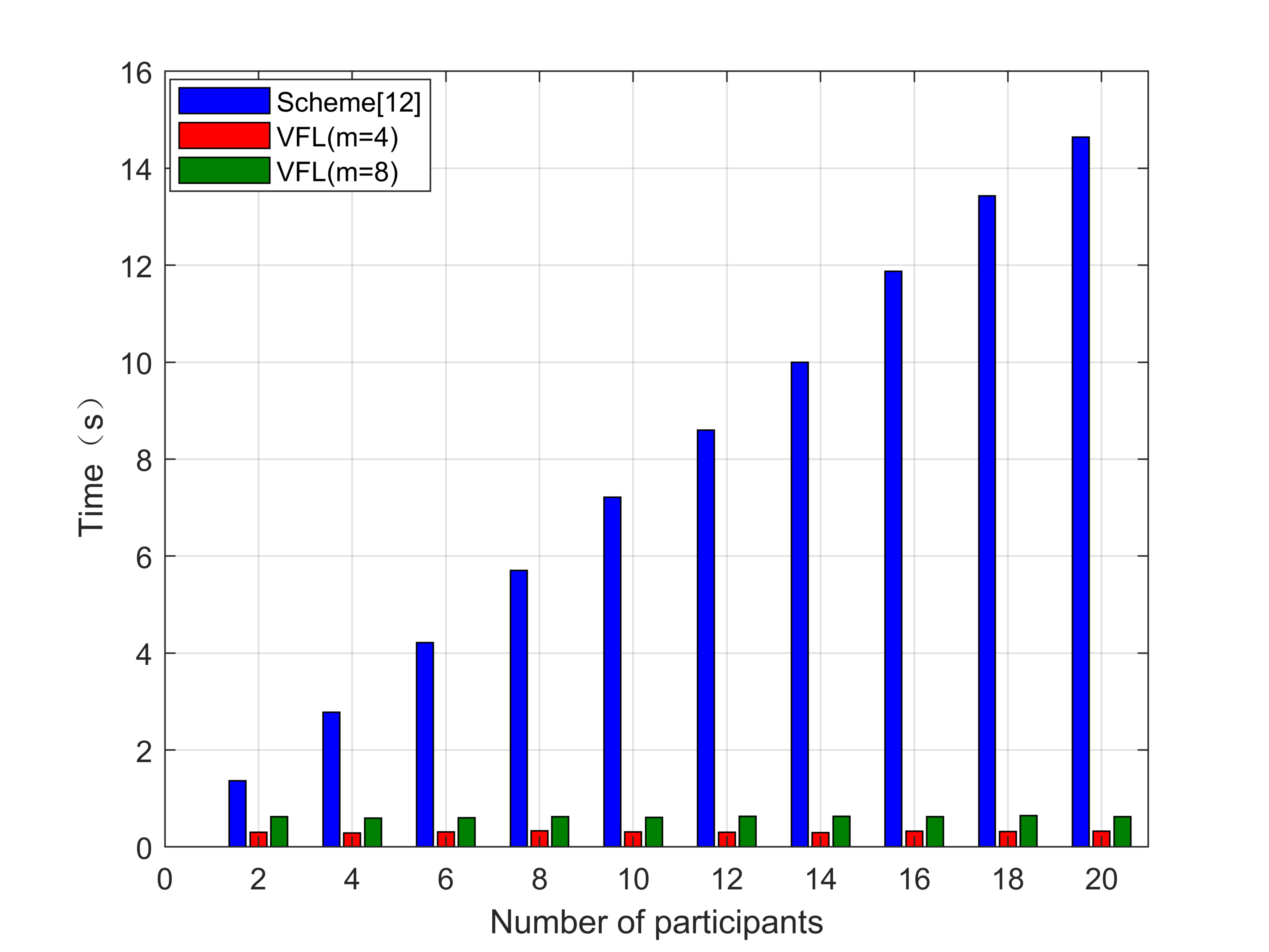}
	\caption{\small{Comparison of verification overhead between different schemes}}
	\label{fig:untitled89}
\end{figure}

\subsection{Communication Overhead}
The communication overhead is related to the size of the uploaded gradient ciphertext and downloaded aggregated ciphertext. Actually, the size of the gradient ciphertext is the same as that of aggregated ciphertext. We discuss communication overhead in terms of communication amount and time. In theory, communication time increases linearly with the communication amount. TABLE I presents the communication amount and time in each round of training for a participant in different schemes. For 64-\textsl{bit} precision, the MLP model gradient parameters are 1192202$\times $64/(8$\times $ $1024^{2}$)$\approx $9.10MB in plain. And our experiment is carried out on a 1Gbps communication channel. Since we use the CRT to reduce the data size, so the communication amount of our VFL is the same as the original federated learning [7], which is about 18.20MB in each interaction, no matter what the value of $m$ is. The participant in Ma's scheme [12] need to upload additional redundant to assist the verification process, the communication amount is approximately twice that of the scheme [7], about 34.16MB. Though the scheme [9] takes the data packaging technology to decrease the encryption operations, its ciphertext size is still about twice that of the scheme [7].
\begin{table}[h]
	\small 
	\caption{\small{C{\upshape ommunication amount and time of different schemes.}}}
	\setlength{\abovecaptionskip}{0.cm}
	\setlength{\belowcaptionskip}{-1cm}
	\begin{center}
		\begin{spacing}{1.3}
			\begin{tabular}{ | m{2.5cm}<{\centering}|m{2.0cm}<{\centering} | m{2.0cm}<{\centering} |}
				\hline
				Scheme & Amount & Time \\ \hline
				Scheme[9] & 34.16MB & 0.272s\\ \hline
				Scheme[12] & 40.34MB & 0.322s\\ \hline
				VFL($m=4$) & 18.20MB & 0.151s\\
				\hline
				VFL($m=8$) & 18.20MB & 0.151s\\
				\hline
			\end{tabular}
		\end{spacing}
	\end{center}
\end{table}

\subsection{Total Overhead}
This section compares the efficiency of different schemes by the total overhead per round. In addition to the computational and communication overhead detailed earlier, the total overhead includes the overhead of training, updating models, and aggregating. In all schemes, participants use the same algorithm to train the model before encryption and update the model after decryption, so they have the same overhead of training and updating model. The aggregation overhead is related to the communication amount and increases with it. We divide the total overhead into three parts: the overhead of a participant $T_{part}$, the aggregation server $T_{ser}$, and communication $T_{com}$. Because participants upload synchronously, the total overhead is $T_{part}$+$T_{ser}$+$T_{com}$. TABLE II presents the summary of the overhead of these schemes for 20 participants to train the MLP model. For our VFL, the total overhead with $m=4$ is 3.053s, and the one with $m=8$ is 10.304s. The total overhead of the scheme [9] is 16.713s, and the scheme [12] is 45.528s. Even if $m=8$, the total overhead of the scheme [9] and [12] are about 1.62 and 4.42 times that of our scheme, respectively. Moreover, it is worth mentioning that the scheme [9] does not have the verification mechanism. Therefore, even with the additional verification, the VFL is still more efficient.
\begin{table}[h]
	\small 
	\caption{\small{T{\upshape he summary of overhead of different schemes.}}}
	\setlength{\abovecaptionskip}{0cm}
	\setlength{\belowcaptionskip}{0cm}
	\begin{center}
		\begin{spacing}{1.3}
			\begin{tabular}{ | m{2.3cm}<{\centering}|m{1.1cm}<{\centering} | m{1.1cm}<{\centering} |m{1.1cm}<{\centering} |m{1.0cm}<{\centering} |}
				\hline
				Scheme & $T_{part}$ & $T_{ser}$ & $T_{com}$ & Total \\ \hline
				Scheme[9] & 16.392s & 0.049s & 0.272s & 16.713s  \\ \hline
				Scheme[12] & 45.149s & 0.057s & 0.322s & 45.528s \\ \hline
				VFL($m=4$) & 2.736s & 0.026s & 0.151s & 3.053s \\
				\hline
				VFL($m=8$) & 9.865s & 0.028s & 0.151s & 10.304s \\
				\hline
			\end{tabular}
		\end{spacing}
	\end{center}
\end{table}

\subsection{Industrial Applications}
With the popularity of artificial intelligence in industrial applications, secure training multi-source data has become a research hotspot in the intelligent industry. The VFL proposed in this paper can be applied to many industrial scenarios.

\begin{figure}[b]
	\centering
	\includegraphics[width=0.9\linewidth]{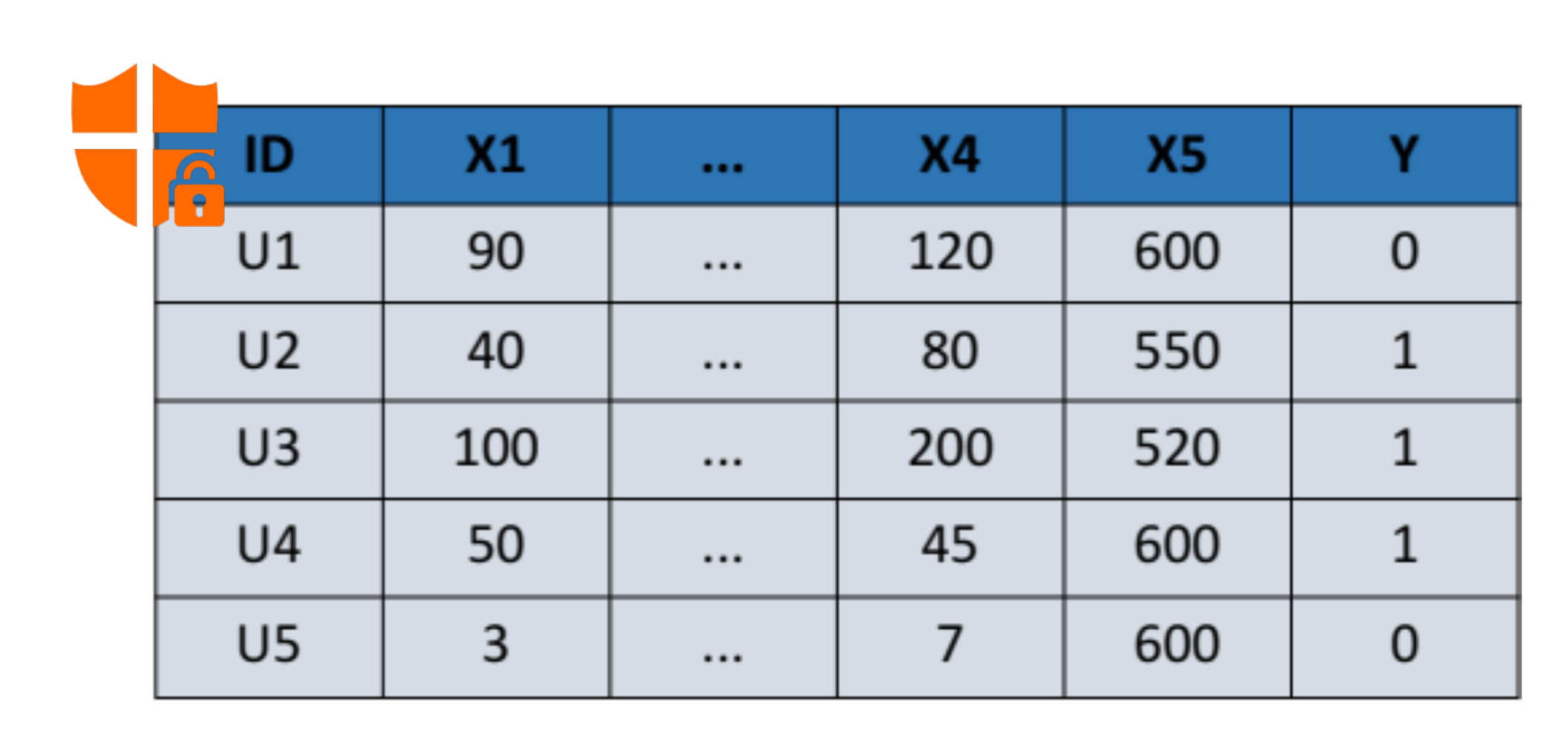}
	\caption{\small{The datasets about enterprises held by a bank.}}
	\label{fig:Dataset}
\end{figure}

\begin{itemize}
	\item[1)]  \textsl{Enterprise Risk Assessment:} With the rapid development of the Internet and the wide application of e-commerce, it has become a research hotspot for banks to establish risk assessment models for enterprises based on their invoice amount and credit data. Fig. 9 shows the datasets about enterprises held by a bank. ID is the taxpayer identity number, $X1$ to $X4$ represent the the invoice amount of the enterprise in each quarter, $X5$ is the credit score and $Y$ indicates whether there is risk (``0'' means the enterprise is in risk.), where $Xi$ is the model input and $Y$ is the label. In reality, the ID held by different banks has a small overlap. Fig. 10 presents the application of VFL in training risk assessment model. Multiple banks efficiently build a reliable and high-quality risk assessment model through collaboration without leaking the enterprise customer data. Moreover, each bank can independently verify whether the third party acting as the aggregation server forges the results to ensure the credibility of the model. On account of there is a competitive relationship between banks, VFL prevents data leakage in the case of collusion.

	\begin{figure}[h]
		\centering
		\includegraphics[width=1.0\linewidth]{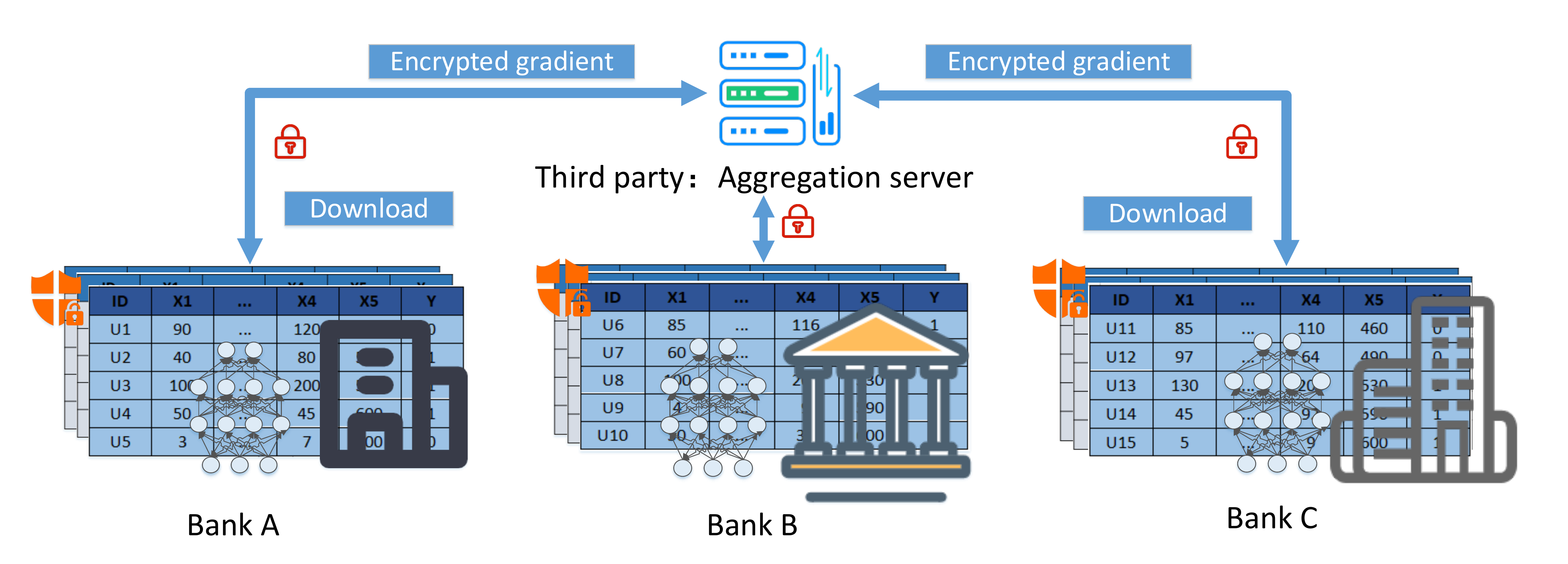}
		\caption{\small{Application of VFL in risk assessment model of multi bank cooperative training}}
		\label{fig:apply}
	\end{figure}

	\item[2)] \textsl{Anti-Money Laundering:} Money laundering samples held by different banks may have different characteristics. VFL framework is able to make use of these samples of several banks to construct a deeper structure model without disclosing the samples. In the process of model training, the input can be the number of large sum transactions, the number of fund sources inconsistent with the business scope and so on. The label is whether it is money laundering.

	\item[3)] \textsl{Medical System:} Due to virus mutation or climate changes, the same disease may show different symptoms in different regions. In addition, with the deepening of medical research, more symptoms will be found. Therefore, the model trained only by local medical data is very likely to be misdiagnosed. The VFL realizes the collaboration of multiple medical institutions to build a high-precision diagnostic model without revealing specific information of patients.
\end{itemize}

In the above industrial scenarios, the protection of the model is equally important, and ownership of the final model is only available to participants in the training. The VFL proposed in this paper avoids the leakage of models and well protects the right of participants.

\section{Conclusions and Future Work}
In this paper, we proposed the VFL, a privacy-preserving and verifiable federated learning framework for industrial intelligent. The VFL framework exploits Lagrange interpolation to carefully set interpolation points, which allows each participant to verify the correctness of the aggregated result effectively. Compared with the exiting verifiable federated learning schemes, the computational overhead of our verification mechanism does not increase with the number of participants. Meanwhile, we take advantage of the blinding technique to protect the trained model and the private gradients of participants. If no more than $\bm{n}$-2 of $\bm{n}$ participants collude with the aggregation server, VFL could guarantee the encrypted gradients of other participants not being inverted. The experiment results on MNIST dataset demonstrated that the VFL provides advantages in terms of verification and total overhead. In the future work, we plan to study more complex neural networks and the data with richer classification labels.

\end{document}